\documentclass[thmsa,12pt]{article}

\usepackage{amssymb}
\usepackage{float}
\usepackage{makeidx}
\usepackage[dvips]{graphicx}
\usepackage{indentfirst}

\title{\bf Extraction of Thermodynamic Data from Ternary Diffusion Coefficients of Lysozyme
Chloride in Water and Aqueous Na$_2$SO$_4$}

\author{Daniela Buzatu\thanks{Physics Department,
Politehnica University, Bucharest, 77206, Romania.}, Emil
Petrescu\footnotemark[1], Florin D. Buzatu\thanks{Department of
Theoretical Physics, National Institute for Physics and Nuclear
Engineering, Bucharest-M\u{a}gurele, 76900, Romania.} \\
John G. Albright\thanks{Department of Chemistry, Texas Christian
University, USA}}

\date{}

\begin{document}

\maketitle

\setlength{\unitlength}{1mm}
\newcommand{\pl}{\partial}
\newcommand{\ep}{\varepsilon}

\begin{abstract}
\noindent This paper presents, for ternary
lysozyme-Na$_2$SO$_4$-water system, the thermodynamic data
extracted from the measured values of four ter\-na\-ry  diffusion
coefficients and the Onsager reciprocal relations. The calculation
for derivatives of solute chemical potentials with respect to
solute molar concentrations was made using the method presented in
\cite{1}. This method is applicable to systems in which the molar
concentration of one solute is very small compared to that of the
other, like in our case. The approach is illustrated for the
lysozyme chloride-Na$_2$SO$_4$-water system at 25$^o$ C, pH 4.5
and at 0.6 mM (8.6 mg/mL) lysozyme chloride and 0.1, 0.25, 0.5,
0.65, and 0.8 M Na$_2$SO$_4$ concentrations. The calculated solute
chemical potential derivatives were used to compute the protein
cation charge approximately. We also compute the diffusion Onsager
coefficients $(L_{ij})_o$ for each composition at pH 4.5.
\end{abstract}

\newpage

\noindent {\bf Motivation} \\

\noindent In order to determine the protein structure through
X-ray diffraction, high quality crystals are required. The protein
crystallization process usually occurs in aqueous solution that
contain salts as precipitant agents. The physical and chemical
properties of these solutions affect drastically the nucleation
and crystal growth processes. Protein aggregation depends on
protein-protein and protein-precipitant interactions in the
solution. In these interactions, the effective charge of the
protein plays an important role. From titration experiments
\cite{2} only the stoichiometric value of the protein charge can
be determine, but it does not take into account the presence of
ions that may bind on the macromolecules and change their net
charge as already shown by diffusion experiments \cite{3}. \\

\noindent The diffusion of protein is one of the fundamental
processes occurring in biological systems, and it is also an
important step in the crystallization mechanism. Crystallization
is an intrinsically non-equilibrium process, and concentration
gradients will occur around the crystal. The protein crystallizes,
reducing its concentration at the moving face of the growing
crystal and creating a protein gradient between the bulk solution
and the crystal. This gradient in turn causes multicomponent
diffusive transport of protein and precipitant. Diffusion in
protein crystal growth inevitably occurs under conditions for
which no species has an uniform concentration raising the issue of
multicomponent diffusion. \\

\noindent The complete description of an $n$-solute system
requires an $n\times n$ matrix of diffusion coefficients relating
the flux of each solute component to the gradients of all solute
components \cite{4}. The importance of other species on protein
diffusion follows from the one-dimensional flux relations
\cite{4}:
\begin{eqnarray}
-J=\sum_{j=1}^n(D_{ij})_v\partial C_j/\partial
x\;\;\;\;\;\;\;i=1,....,n
\end{eqnarray}
in which the cross-term diffusion coefficients (off-diagonal
elements $(D_{ij})_v\;i\neq j$) can be positive or negative. In
ternary systems ($n=2$), our case, the one-dimensional flux
relations could be written as:
\begin{eqnarray}
-J_1 & = & (D_{11})_v\frac{\partial C_1}{\partial
x}+(D_{12})_v\frac{\partial
C_2}{\partial x} \\
-J_2 & = & (D_{21})_v\frac{\partial C_1}{\partial
x}+(D_{22})_v\frac{\partial C_2}{\partial x}
\end{eqnarray}
where $J_1$ and $J_2$ - the protein flux and respectively salt
flux, $(D_{11})_v$ and $(D_{22})_v$ - the main-term diffusion
coefficients relating to the flux of component to its own
concentration gradient, and $(D_{12})_v$ and $(D_{21})_v$ - the
cross-term diffusion coefficients relating the flux of each
component to the gradient of the other. The index $v$ from the
diffusion coefficients shows that the experiment were done under
the assumption the volume change on mixing and changes in
concentrations across the diffusion boundary were small.
Consequently, with a good approximation, the measured diffusion
coefficients may be considered to be for the volume-fixed
reference frame \cite{5} defined by:
\begin{eqnarray}
\sum_{i=0}^nJ_i\bar{V}_i=0
\end{eqnarray}
where $\bar{V}_i$ is the partial molar volume of the $i$th
species, and the subscript 0 denotes the solvent. \\
The importance of multicomponent diffusion has been recognized in
the crystal growth community \cite{6,7} and a crystal growth model
has properly accounted for multicomponent diffusive transport in
lysozyme chloride-NaCl-water system \cite{8,1}. The experimental
multicomponent diffusion coefficients are essential for accurate
modeling of protein transport, especially in view of the very
large cross-term coefficient $(D_{21})_v$ reported here. Moreover,
the concentration of supporting electrolyte dependence of all the
diffusion coefficients should be important for supersaturation
region and also for its directly contribution to the protein flux.
\\
The use of ternary diffusion coefficients to determine binding
coefficients and other thermodynamic data is very well established
\cite{9,10} . For our ternary system, the molar concentration of
one solute is very small compared to that of the other, and also
small enough that an inverse concentration dependence dominates
certain activity coefficient derivatives. For such systems, using
the Onsager reciprocal relations (ORR), along with precision
measurements of ternary diffusion coefficients from our earlier
paper \cite{11}, we determined concentration derivatives of the
chemical potentials ($\mu_{ij}\equiv \partial\mu_i/\partial C_j$)
of two solutes with respect to one solute molar concentrations. \\
\noindent In our ternary experiments, the molarity of lysozyme
chloride is small compared to that of Na$_2$SO$_4$. Thus, the
self-derivative for lysozyme chloride, $\mu_{11}$, is dominated by
its concentration term. The self-derivative $\mu_{22}$ for
Na$_2$SO$_4$ is essentially that of the binary with minor
correction. In order to obtain the molarity cross-derivatives
($\mu_{12}$ and $\mu_{21}$ which are unequal \cite{12}) we have to
use two additional equations: (1) equality of the molality
cross-derivatives \cite{13}:
\begin{eqnarray}
\frac{\partial\mu_1}{\partial m_2}=\frac{\partial\mu_2}{\partial
m_1}
\end{eqnarray}
where $m_i$ is the molality of solute $i$, and (2) the ORR
equation:
\begin{eqnarray}
\mu_{11}(D_{12})_o-\mu_{12}(D_{21})_o=\mu_{22}(D_{21})_o-\mu_{21}(D_{22})_o
\end{eqnarray}
relating the four molarity derivatives and the ternary difussion
coefficients in a solvent-fixed reference frame $(D_{ij})_o$
\cite{14,15}. This method \cite{1} yields to an estimate of
lysozyme charge and a functional approximation to the change of
the chemical potential of lysozyme chloride with Na$_2$SO$_4$
concentration. This, together with the diffusion coefficients,
will permit the modeling of protein crystallization processes. \\

\noindent {\bf Experimental section} \\

\noindent All the experimental work was performed al Texas
Christian University, in the Chemistry Department. \\ \\
\noindent {\bf Materials}. All the materials, solution preparation
procedures, apparatus and density measurement procedures are
described in the work \cite{8}. We used a hen egg-white lysozyme,
recrystallized six times purchased from Seikagaku America.  \\
The molecular mass of the lysozyme solute, $M_1$, was taken as
14307 g/mol, and this value \cite{16} was used to calculate all
concentrations after correction for the moisture and chloride
content. Buoyancy corrections were made with the commonly used
lysozyme crystal density \cite{17,18,19} of 1.305 g/cm$^3$.\\
The molecular mass of water, $M_o$, was taken as 18.015 g/cm$^3$
and the molecular mass of Na$_2$SO$_4$, M$_2$, was taken as
142.037 g/mol. \\
\noindent Mallinckrodt reagent HCl ($\sim$ 12 M) was diluted by
half with pure water and distilled at the constant boiling
composition. This resulting HCl solution ($\sim $ 6 M) was then
diluted (pH 1.2) and used to adjust the
pH of solution.\\

\noindent {\bf Preparation of Solutions}. All solutions were
prepared by mass with appropriate buoyancy corrections. All
weighings were performed with a Mettler Toledo AT400
electrobalance. Since the as-received lysozyme powder was very
hygroscopic, all manipulations in which water absorption might be
critical were performed in a dry glove box. Stock solutions of
lysozyme were made by adding as-received protein to a pre-weighted
bottle that had contained dry box air, capping the bottle, and
reweighing to get the weight and thus mass of lysozyme. Water was
added to dissolve the lysozyme, and the solution was weighed. An
accurate density measurement was made and used to obtain the
molarity of the stock solution. \\
\noindent For ternary experiments, precise masses of Na$_2$SO$_4$
were added to flasks containing previously weighed quantities of
lysozyme stock solutions. These solutions were mixed and diluted
to within 10 cm$^3$ of the final volume. The pH was adjusted, and
the solutions were diluted to their final mass. \\ \\
{\bf Measurements of pH}. The pH measurements were made using a
Corning model 130 pH meter with an Orion model 8102 combination
ROSS pH electrode. The meter was calibrated with standard pH 7 and
pH 4 buffers and checked against a pH 5 standard buffer.
\\ \\
{\bf Density Measurements}. All density measurements were made
with a Mettler-Paar DMA40 density meter, with an RS-232 output to
a Apple $\Pi$+. By time averaging the output, a precision of
0.00001 g/cm$^3$ or better could be achieved. The temperature of
the vibrating tube in the density meter was controlled with water
from a large well-regulated water bath whose temperature was
25.00$\pm$ 0.01 $^o$ C.
 \\ \\
{\bf Free-Diffusion  Measurements}. For binary Na$_2$SO$_4$-water
and ternary Lys-Na$_2$SO$_4$-Water we performed measurements for
free-diffusion using the high-precision Gosting diffusiometer
\cite{20,21,22} operated in its Rayleigh interferometric optical
mode. The procedure for measuring binary (D$_2)_v$ and ternary
diffusion coefficients (D$_{ij})_v$ were described in detail in
the work \cite{8}. In order to measure the four diffusion
coefficients of the system, the experiments must be performed with
at least two different concentration differences at each
combination of mean concentration \cite{20,23,24}. For ternary
experiments, for each pair of mean concentrations, two
measurements were performed with $\alpha_1=0$ and the two with
$\alpha_1=0.8$ ($\alpha_i$ - the
refractive index fraction due to the $i$th solute \cite{8}). \\
\noindent In order to make the data analysis of the free-diffusion
experiments we used the Fick's second law:
\begin{eqnarray}
\frac{\partial C_i}{\partial t}=\sum_{j=1}^2
D_{ij}\frac{\partial^2C_j}{\partial x^2}\;\;\;\;\;\;\;i=1,2
\end{eqnarray}
for two solutes. We made the assumption that the concentration
differences of the solutes across the initial boundary are small
enough and the diffusion coefficients are constant \cite{25}. Also
the volume changes on mixing were negligible, thus all the
measured diffusion coefficients are given relative to the
volume-fixed frame of reference defined by equation (4). Also we
could made a comparison with the dynamic light scattering
\cite{27} results from our previous paper \cite{11}. \\

\noindent {\bf Results} \\

\noindent {\bf Ternary diffusion experiments} were performed on
the lysozyme chloride-Na$_2$SO$_4$-water system at pH$=4.5$ and
T=$25^0$ C. The four ternary diffusion coefficients obtained,
(D$_{11})_v$, (D$_{22})_v$, (D$_{12})_v$ and (D$_{21})_v$ were
published in our earlier paper \cite{11} for a mean Na$_2$SO$_4$
concentration of 0.1, 0.25, 0.5, 0.65 and 0.8 M. The
interferometric data for the diffusion coefficients (D$_{11})_v$,
(D$_{22})_v$, (D$_{12})_v$ and (D$_{21})_v$ are reported in the
Tables 1,2,3. \\

\noindent {\bf Partial molar volumes} values, $\bar{V}_1$,
$\bar{V}_2$ and $\bar{V}_o$, were calculated for each component
using eqs A-7 ($q=2$) and 5 in \cite{14} and reported in the Tables 1,2,3. \\
\noindent Values of mean density $\bar{d}$ and $H_i=(\partial
d/\partial C_i)_{T,p,C_j,j\neq i}$ in the Tables 1,2,3 were
calculated using densities of all eight solutions from each
experiment set. Densities were assumed to be linear in solute
concentrations respecting the equation \cite{8}:
\begin{eqnarray}
d=\bar d+H_1(C_1-\bar{\bar{C}_1})+H_2(C_2-\bar{\bar{C}_2})
\end{eqnarray}
where $\bar{\bar{C}_1}$ and $\bar{\bar{C}_2}$ are the averages of
the mean concentrations for all four experiments in a series. \\

\noindent {\bf Use of irreversible thermodynamics and diffusion
data to calculate chemical potential derivatives}. For our ternary
system lysozyme chloride-Na$_2$SO$_4$, the molar concentration of
a macromolecule and the supporting electrolyte are small and
large, respectively. So, in this case it is possible to estimate
the chemical potential derivatives
$\mu_{11}$ and $\mu_{22}$ \cite{1}. \\
\noindent The molality cross-derivative relation, eq.(5), comes
from classical thermodynamics. From eq. (5), an expression can be
derived relating the four molarity partial derivatives $\mu_{ij}$
\cite{26}. The eq.(6), the ternary ORR of irreversible
thermodynamics, relates the four $(D_{ij})_o$ values of diffusion
coefficients in a solvent-fixed reference frame and the four
$\mu_{ij}$ values \cite{14,15}. \\

\noindent {\bf Fundamental equations}. The analysis of chemical
potential derivatives was made in terms of quantities referred to
a solvent-fixed reference frame, identify by a subscript $0$, with
the diffusion Onsager coefficients denoted by $(L_{ij})_o$. The
solvent-fixed $(D_{ij})_o$ values shown in the Tables 1,2,3 were
obtained from experimental volume-fixed $(D_{ij})_v$ values by
standard equations involving the $\bar{V}_i$ \cite{1,18,19}.\\
\noindent The diffusion Onsager coefficients $(L_{ij})_o$ and the
solvent-fixed $(D_{ij})_o$ diffusion coefficients are related (in
matrix form) by:
\begin{eqnarray}
\left[\begin{array}{cc}
(D_{11})_o & (D_{12})_o \\
(D_{21})_o & (D_{22})_o
\end{array}\right]= \nonumber \\
\left[\begin{array}{cc} (L_{11})_o\mu_{11}+(L_{12})_o\mu_{21} &
(L_{11})_o\mu_{12}+(L_{12})_o\mu_{22} \\
(L_{21})_o\mu_{11}+(L_{22})_o\mu_{21} &
(L_{21})_o\mu_{21}+(L_{22})_o\mu_{22}
\end{array}\right]
\end{eqnarray}

The inverse realtion is:
\begin{eqnarray}
\left[\begin{array}{cc}
(L_{11})_o & (L_{12})_o \\
(L_{21})_o & (L_{22})_o
\end{array}\right]=
\frac{1}{\mu_{11}\mu_{22}-\mu_{12}\mu_{21}}\times \nonumber \\
\left[\begin{array}{cc} \mu_{22}(D_{11})_o-\mu_{21}(D_{12})_o &
\mu_{11}(D_{12})_o-\mu_{12}(D_{11})_o \\
\mu_{22}(D_{21})_o-\mu_{21}(D_{22})_o &
\mu_{11}(D_{22})_o-\mu_{12}(D_{21})_o
\end{array}\right]
\end{eqnarray}
Since the ORR, $(L_{12})_o=(L{_21})_o$ \cite{14,15} apply to the
solvent-fixed frame, eq. (10) yields eq. (6). \\
\noindent Also, from eq. (5) for cross-derivative molality and the
relations between $C_i$ and $m_i$, we can show that \cite{11}:
\begin{eqnarray}
\frac{1}{C_oM_o}\frac{\partial\mu_1}{\partial
m_2}=\mu_{12}(1-C_2\bar{V}_2)-\mu_{11}C_1\bar{V}_2= \\ \nonumber
\mu_{21}(1-C_1\bar{V}_1)-\mu_{22}C_2\bar{V}_1=\frac{1}{C_oM_o}\frac{\partial\mu_2}{\partial
m_1}
\end{eqnarray}
where $M_o$ is the molecular mass of water. \\
\noindent In order to calculate the $\mu_{ij}$ molarity
cross-derivatives, we took in account their general thermodynamics
expressions in terms of volume concentrations and the
corresponding mean ionic activity coefficients $y_i$ for volume
concentrations \cite{26}, and also we assumed that lysozyme
chloride has stoichiometry LyCl$_{z_p}$. Thus, the $\mu_{ij}$ for
our case could be written in a matrix form as:
\begin{eqnarray}
\left[\begin{array}{cc}
\mu_{11} & \mu_{12} \\
\mu_{21} & \mu_{22}
\end{array}\right]=RT\times \\ \nonumber
\left[\begin{array}{cc}
\frac{1}{C_1}+\frac{z_p^2}{z_p/2C_1+C_2}+(z_p+1)\frac{\partial\ln{y}_1}{\partial
C_1} &
\frac{z_p}{z_p/2C_1+C_2}+(z_p+1)\frac{\partial\ln{y}_1}{\partial
C_2} \\
\frac{z_p}{z_p/2C_1+C_2}+2\frac{\partial\ln{y}_2}{\partial C_1} &
\frac{1}{C_2}+\frac{1}{z_p/2C_1+C_2}+2\frac{\partial\ln{y}_2}{\partial
C_2}
\end{array}\right]
\end{eqnarray}
where the quantity $z_p/2C_1+C_2$ is equivalent to the total
normality $N$ of our ternary solution. \\
\noindent Using the eq.(12) we computed the partial derivatives of
chemical potentials, $\mu_{ij}$, for the case in which the
molarity of at least one component is very low.\\
\noindent From eq.(12) we calculated the $\mu_{11}$ and $\mu_{22}$
taking in account the terms which are dominant for our case. In
order to calculate the other two derivatives chemical potentials,
$\mu_{12}$ and $\mu_{21}$ we use the eqs. (6) and (11) and the
four $(D_{ij})_o$:
\begin{eqnarray}
\mu_{12}=\{\mu_{11}[C_1\bar{V}_2(D_{22})_o-(1-C_1\bar{V}_1)(D_{12})_o]-
\nonumber \\
\mu_{22}[C_2\bar{V}_1(D_{22})_o-(1-C_1\bar{V}_1)(D_{21})_o]\}/
\nonumber \\
\{(1-C_2\bar{V}_2(D_{22})_o-(1-C_1\bar{V}_1)(D_{11})_o\} \\
\mu_{21}=\{\mu_{11}[C_1\bar{V}_2(D_{11})_o-(1-C_2\bar{V}_2)(D_{12})_o]-
\nonumber \\
\mu_{22}[C_2\bar{V}_1(D_{11})_o-(1-C_2\bar{V}_2)(D_{21})_o]\}/
\nonumber \\
\{(1-C_2\bar{V}_2(D_{22})_o-(1-C_1\bar{V}_1)(D_{11})_o\}
\end{eqnarray}
All the values for derivative chemical potentials $\mu_{ij}$ are
reported in the Tables 1,2,3 for each salt concentration.\\
\noindent Using the eqs. (10), (12), (13) and (14) we calculated
the thermodynamics transport coefficients $(L_{ij})_o$ for pH 4.5
and $C_1=0.60$ mM, which are reported also in the Tables 1,2,3. \\

\noindent {\bf Calculation of lysozyme chloride charge using the
values for $\mu_{12}$ and $\mu_{21}$} is described in \cite{1}.
Looking into eq. (12), the expressions for $\mu_{12}$ and
$\mu_{21}$ suggest that we can obtain the lysozyme chloride charge
multiplying $\mu_{12}/RT$ and $\mu{_21}/RT$ by $z_p/2C_1+C_2$ and
we'll receive the following dependence:
\begin{eqnarray}
Y_{12}=(z_p/2C_1+C_2)\frac{\mu_{12}}{RT} & = &
z_p+(z_p/2C_1+C_2)(z_p+1)\frac{\partial\ln{y}_1}{\partial C_2} \\
Y_{21}=(z_p/2C_1+C_2)\frac{\mu_{21}}{RT} & = &
z_p+2(z_p/2C_1+C_2)\frac{\partial\ln{y}_2}{\partial C_1}
\end{eqnarray}
The dependence of $Y_{21}$ and $Y_{21}$ on $(z_p/2C_1+C_2)$ is
shown in the Fig.1. From the graph we could estimate the protein
charge, $z_p$ and to compare it with the value obtain from
thermodynamics data from \cite{1}. At pH 4.5 ,the average value of
$z_p$ obtained from eqs. (15) and (16) is $z_p=3.29$, in
comparison with the value obtained, using the same approach, for
lysozyme-NaCl-Water system, $z_p=8.9$. This data were obtained
from thermodynamics data for ternary system, which were in turn
obtained in part from transport data. \\

\noindent {\bf Conclusions} We reported the complete set of
multicomponent diffusion coefficients for ternary
lys-Na$_2$SO$_4$-water system at concentrations high enough to be
relevant to crystallization studies, in the volume-fixed frame,
$(D_{ij})_v$ and in the solvent-fixed frame, $(D_{ij})_o$. Also we
calculated the derivatives chemical potentials $\mu_{ij}$ for our
ternary system and after that we estimated the lysozyme chloride
charge, $z_p$, from irreversible thermodynamics and diffusion
data. We also reported the four thermodynamics transport
coefficients $(L_{ij})_o$ at pH 4.5 and $C_1=0.60$ mM.
\\ \\
\noindent {\bf Acknowledgment.} One of the authors (DB), is very
grateful to O. Annunziata for constant and helpful advices during
the experimental work. This research was supported by the Texas
Christian University Grant RCAF-11950 and by the NASA Microgravity
Biotechnology Program through the Grant NAG8-1356.

\begin{figure}
\hspace{0.7cm}
\includegraphics[scale=1]{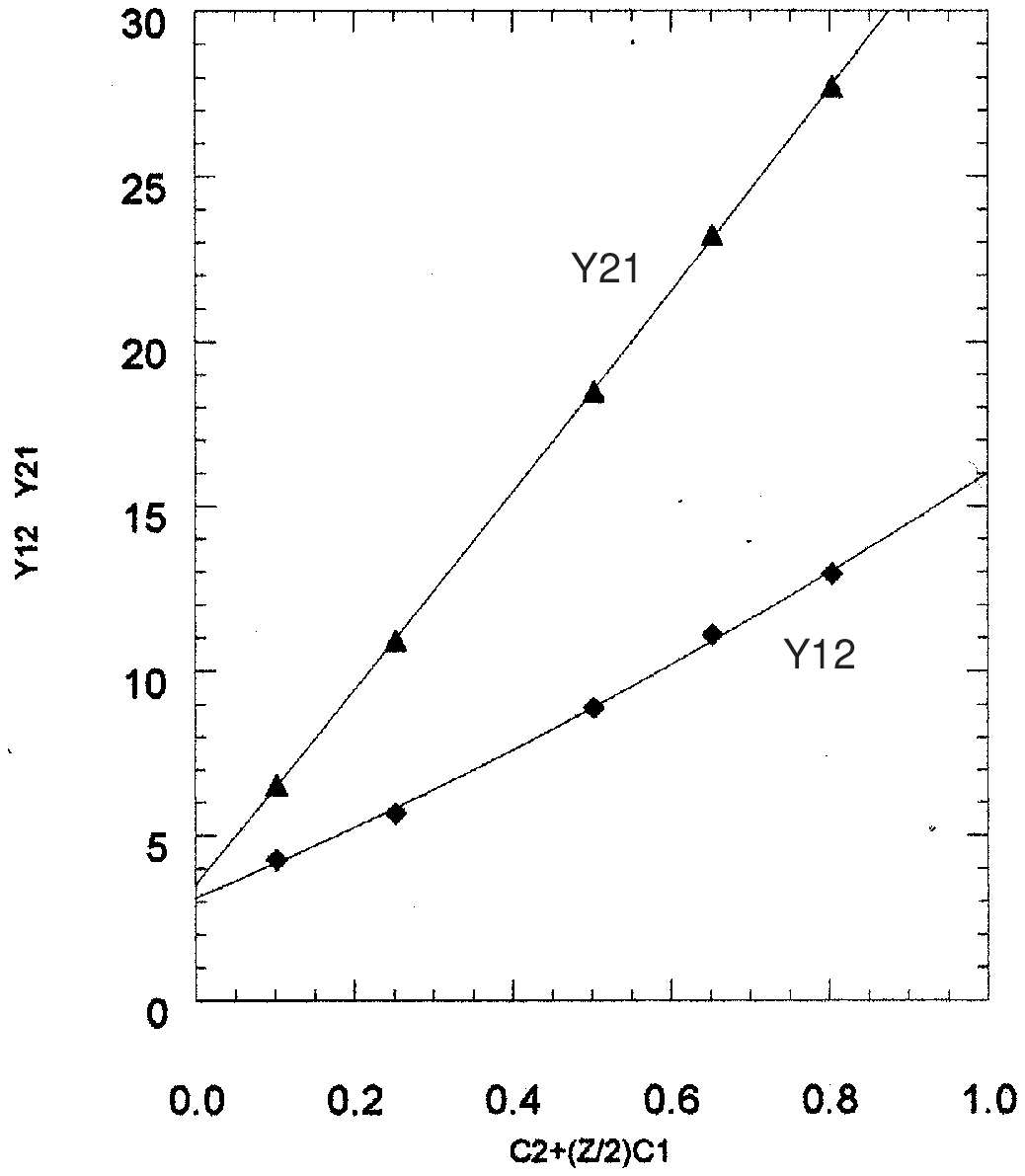}
\begin{center}
{\bf Fig. 1}
\end{center}
\end{figure}

\begin{table}[h]
\begin{center}
{\bf Table 1} \\ \vspace{0.3cm}
\begin{tabular}{lcc}
$\bar{\bar{C}_1}$(mM)  & 0.6000 & 0.6000
\\
$\bar{\bar{C}_2}$(M) & 0.1000 & 0.2500  \\
$\bar{d}(g cm^{-3})$ & 1.012189 & 1.030671  \\
$H_1(10^3 g mol^{-1}$ & 4.286 & 1.149  \\
$H_2(10^3 g mol^{-1}$ & 0.12500 & 0.12230  \\
$\bar{V}_1(cm^3 mol^{-1})$ & 10050 & 10182 \\
$\bar{V}_2(cm^3 mol^{-1}) $& 17.090 & 19.780 \\
$\bar{V}_o(cm^3 mol^{-1})$ & 18.067 & 18.058 \\
$(D_{11})_v(10^{-9}m^2s^{-1})$ & 0.1169$\pm$ 0.0001 & 0.1090$\pm$
0.0001 \\
$(D_{12})_v(10^{-9}m^2s^{-1})$ & -0.000013$\pm$ 0.000001 &
0.000108$\pm$ 0.000001  \\
$(D_{21})_v(10^{-9}m^2s^{-1})$& 2.49$\pm$ 0.01 & 4.14$\pm$ 0.01 \\
$(D_{22})_v(10^{-9}m^2s^{-1})$ & 0.9661$\pm$ 0.0001 & 0.8826$\pm$
0.0001 \\
$(D_{11})_o(10^{-9}m^2s^{-1})$ & 0.1176 & 0.1097 \\
$(D_{12})_o(10^{-9}m^2s^{-1})$ & -0.0000031 & 0.0001193 \\
$(D_{21})_o(10^{-9}m^2s^{-1}) $ & 2.6 & 4.5 \\
$(D_{22})_o(10^{-9}m^2s^{-1})$ & 0.968 & 0.887 \\
$\mu_{11}/RT(M^{-1})$ & 1787 & 1715 \\
$\mu_{12}/RT(M^{-1})$  & 41.9 & 22.5 \\
$\mu_{21}/RT(M^{-1}) $ & 64.2 & 43.3 \\
$\mu_{22}/RT(M^{-1})$ & 21.894 & 8.131 \\
$RT(L_{11})_o(10^{-9}Mm^2s^{-1})$ & 0.0000707 & 0.0000684\\
$RT(L_{12})_o(10^{-9}Mm^2s^{-1})$ & -0.0001354 & -0.0017452 \\
$RT(L_{22})_o(10^{-9}Mm^2s^{-1})$ & 0.044487 & 0.1094762
\end{tabular}
\end{center}
\end{table}

\begin{table}[h]
\begin{center}
{\bf Table 2} \\ \vspace{0.3cm}
\begin{tabular}{lcc}
$\bar{\bar{C}_1}$(mM)  & 0.6000 & 0.6000
\\
$\bar{\bar{C}_2}$(M) & 0.5000 & 0.6500  \\
$\bar{d}(g cm^{-3})$ &  1.060538 & 1.078032
 \\
$H_1(10^3 g mol^{-1}$ & 4.104 & 4.049  \\
$H_2(10^3 g mol^{-1}$ &  0.11733 & 0.11610
 \\
$\bar{V}_1(cm^3 mol^{-1})$ & 10209 & 10257 \\
$\bar{V}_2(cm^3 mol^{-1}) $& 24.720 & 25.930
 \\
$\bar{V}_o(cm^3 mol^{-1})$ &  18.026 & 18.013
\\
$(D_{11})_v(10^{-9}m^2s^{-1})$
& 0.0969$\pm$ 0.0001 & 0.0894$\pm$ 0.0001 \\
$(D_{12})_v(10^{-9}m^2s^{-1})$ & 0.000132$\pm$ 0.000001 &
0.000134$\pm$ 0.000001 \\
$(D_{21})_v(10^{-9}m^2s^{-1})$ & 6.75$\pm$
0.01 & 8.26 $\pm$ 0.01  \\
$(D_{22})_v(10^{-9}m^2s^{-1})$ & 0.7791 $\pm$ 0.0001 & 0.7294 $\pm$ 0.0001 \\
$(D_{11})_o(10^{-9}m^2s^{-1})$ & 0.0977& 0.0902 \\
$(D_{12})_o(10^{-9}m^2s^{-1})$ & 0.0001444 & 0.0001475 \\
$(D_{21})_o(10^{-9}m^2s^{-1}) $ & 7.3 & 9.0 \\
$(D_{22})_o(10^{-9}m^2s^{-1})$ & 0.790 & 0.742 \\
$\mu_{11}/RT(M^{-1})$ & 1691 & 1685 \\
$\mu_{12}/RT(M^{-1})$  & 17.7 & 17 \\
$\mu_{21}/RT(M^{-1}) $ & 36.8& 35.6\\
$\mu_{22}/RT(M^{-1})$ & 3.755 & 2.811 \\
$RT(L_{11})_o(10^{-9}Mm^2s^{-1})$ & 0.0000634 & 0.0000601 \\
$RT(L_{12})_o(10^{-9}Mm^2s^{-1})$ & -0.0002606 & -0.0003110 \\
$RT(L_{22})_o(10^{-9}Mm^2s^{-1})$ & 0.2117597 & 0.31738664
\end{tabular}
\end{center}
\end{table}

\begin{table}[h]
\begin{center}
{\bf Table 3} \\ \vspace{0.3cm}
\begin{tabular}{lc}
$\bar{\bar{C}_1}$(mM)  & 0.6000
\\
$\bar{\bar{C}_2}$(M) &  0.8000 \\
$\bar{d}(g cm^{-3})$ &  1.095745 \\
$H_1(10^3 g mol^{-1}$ &  3.954 \\
$H_2(10^3 g mol^{-1}$ & 0.11502 \\
$\bar{V}_1(cm^3 mol^{-1})$ &  10341\\
$\bar{V}_2(cm^3 mol^{-1}) $& 26.980 \\
$\bar{V}_o(cm^3 mol^{-1})$ & 17.993 \\
$(D_{11})_v(10^{-9}m^2s^{-1})$ & 0.0822$\pm$ 0.0001 \\
$(D_{12})_v(10^{-9}m^2s^{-1})$ & 0.000130$\pm$ 0.000001 \\
$(D_{21})_v(10^{-9}m^2s^{-1})$& 9.50$\pm$ 0.01 \\
$(D_{22})_v(10^{-9}m^2s^{-1})$ & 0.6900$\pm$ 0.0001 \\
$(D_{11})_o(10^{-9}m^2s^{-1})$ & 0.0829 \\
$(D_{12})_o(10^{-9}m^2s^{-1})$ & 0.0001418 \\
$(D_{21})_o(10^{-9}m^2s^{-1}) $ & 10.4 \\
$(D_{22})_o(10^{-9}m^2s^{-1})$ & 0.706\\
$\mu_{11}/RT(M^{-1})$ & 1682 \\
$\mu_{12}/RT(M^{-1})$  & 16.1\\
$\mu_{21}/RT(M^{-1}) $ & 34.5 \\
$\mu_{22}/RT(M^{-1})$ & 2.241 \\
$RT(L_{11})_o(10^{-9}Mm^2s^{-1})$ & 0.1808868 \\
$RT(L_{12})_o(10^{-9}Mm^2s^{-1})$ & -0.000341 \\
$RT(L_{22})_o(10^{-9}Mm^2s^{-1})$ & 0.3173864
\end{tabular}
\end{center}
\end{table}

\end{document}